\documentclass[prd,twocolumn,showpacs,preprintnumbers,amsmath,amssymb,superscriptaddress,nofootinbib,english]{revtex4-2}

\usepackage{graphicx}
\usepackage{dcolumn}
\usepackage{bm}
\usepackage{epsfig}
\usepackage{graphicx}
\usepackage{hyperref}
\usepackage[usenames]{color}
\usepackage{url}
\usepackage{enumitem}
\usepackage{float}
\usepackage[utf8]{inputenc}
\usepackage{cancel}

\usepackage{multirow}
\usepackage{array}
\usepackage{subfloat}
\usepackage{placeins}
\usepackage{subfigure}
\usepackage[normalem]{ulem}
\usepackage[dvipsnames]{xcolor}

\newcommand\redsout{\bgroup\markoverwith{\textcolor{red}{\rule[0.5ex]{2pt}{0.5pt}}}\ULon}

\hypersetup{
    colorlinks=true,
    linkcolor=blue,
    citecolor=blue,
}

\newcommand{\remove}[1]{}

\def\be{\begin{equation}}
\def\ee{\end{equation}}
\def\ba{\begin{eqnarray}}
\def\ea{\end{eqnarray}}

\def\rd{r_{\rm d}}
\def\rdh{r_{\rm d}h}
\def\om{\Omega_{\rm m}h^2}
\def\Om{\Omega_{\rm m}}

\begin{document}

\title{Is Dynamical Dark Energy Necessary? DESI BAO and Modified Recombination}

\author{Seyed Hamidreza Mirpoorian}
\email[]{smirpoor@sfu.ca}
\affiliation{Department of Physics, Simon Fraser University, Burnaby, British Columbia, Canada V5A 1S6}

\author{Karsten Jedamzik}
\email[]{karsten.jedamzik@umontpellier.fr}
\affiliation{Laboratoire de Univers et Particules de Montpellier, UMR5299-CNRS, Universite de Montpellier, 34095 Montpellier, France}

\author{Levon Pogosian}
\email[]{levon@sfu.ca}
\affiliation{Department of Physics, Simon Fraser University, Burnaby, British Columbia, Canada V5A 1S6}

\begin{abstract}
Recent measurements of baryon acoustic oscillations (BAO) by the Dark Energy Spectroscopic Instrument (DESI) exhibit a mild-to-moderate tension with cosmic microwave background (CMB) and Type Ia supernova (SN) observations when interpreted within the $\Lambda$CDM framework. This discrepancy has been cited as evidence for dynamical dark energy (DDE). Given the profound implications of DDE for fundamental physics, we explore whether the tension can instead be resolved by modifying the physics of recombination. We find that a phenomenological model of modified recombination can effectively reconcile the BAO and CMB datasets and, unlike DDE, also predicts a higher Hubble constant $H_0$, thereby partially alleviating the Hubble tension. A global fit to BAO, CMB, and calibrated SN data favors modified recombination over DDE.

\end{abstract}

\maketitle

\section{Motivation and Context}

The $\Lambda$CDM model has been remarkably successful in describing a wide range of cosmological observations, including the anisotropies in the cosmic microwave background (CMB) and the large-scale structure of the Universe. With the advent of increasingly precise observational surveys, the model continues to be tested with growing rigor. While it appears unlikely that $\Lambda$CDM is far from the correct description of the Universe, its two defining components, cold dark matter (CDM) and the cosmological constant $\Lambda$, remain poorly understood at a fundamental level. It is therefore plausible that the model may need to be modified or extended with additional physical ingredients.

The Dark Energy Spectroscopic Instrument (DESI) has recently released its one-year results~\cite{DESI:2024mwx} (DR1) and three-year results~\cite{DESI:2025zgx} (DR2), providing precise measurements of the $\sim 145$ Mpc comoving acoustic scale in the distribution of galaxies, commonly known as baryon acoustic oscillations (BAO). This scale is also imprinted in the cosmic microwave background (CMB) as the Doppler peaks in its angular power spectrum\footnote{More precisely, the location of the first Doppler peak corresponds to the comoving sound horizon at the maximum of the visibility function, $r_\star$, while the BAO scale corresponds to the sound horizon at the epoch of baryon decoupling, or the ``drag'' epoch, $\rd$. The two are closely related, with $\rd \approx 1.02 r_\star$ in $\Lambda$CDM and nearly all its tested extensions (see also the tables in Appendix A).}. By observing a large number of galaxies of different types, along with quasars and the Lyman-$\alpha$ forest, DESI has mapped the BAO feature across multiple redshifts in the range $z \in [0.3, 2.3]$, achieving significantly improved precision over previous surveys~\cite{eBOSS:2020yzd}. Depending on whether it is measured perpendicular to the line of sight, parallel to it, or through the angle-averaged method~\cite{Eisenstein:2005su}, the BAO feature observed at a given redshift $z$ constrains quantities $D_M(z)/r_{\rm d}$, $H(z)r_{\rm d}$ and $D_V(z)/r_{\rm d}$, respectively, where $r_{\rm d}$ is the comoving sound horizon at the drag epoch, $H(z)$ is the Hubble rate, $D_M(z)$ is the comoving angular diameter distance, and $D_V(z) = \left[ czD_M^2(z)/H(z) \right]^{1/3}$. Measurements across multiple redshifts therefore allow one to track the evolution of the comoving distance, yielding constraints on the matter density and possible deviations from a cosmological constant.

The DESI DR2 BAO data alone, in the $\Lambda$CDM model, provides a precise determination of the product of $\rd$ and the dimensionless Hubble parameter, $h = H_0/(100 \ {\rm km/s/Mpc})$, and the matter density parameter $\Om$~\cite{DESI:2025zgx}:
\begin{eqnarray}
\rd h & = & 101.54 \pm 0.73 ~{\rm Mpc}, \nonumber \\
\Om & = & 0.2975 \pm 0.0086 .
\label{eq:DR2}
\end{eqnarray}
This inference does not require knowing $\rd$ and, therefore, is independent of recombination-era physics. The DESI DR2 values are in good agreement with DESI DR1, with smaller error bars. The same parameters can also be tightly constrained using the CMB data, though such an inference depends not only on the late-time expansion history but also on the physics of recombination. Within the $\Lambda$CDM framework, Planck measurements (in this paper we use the combination of the NPIPE {\tt CamSpec} high-$\ell$ likelihood, PR4 CMB lensing, and PR3 low-$\ell$ TT and EE)~\cite{Planck:2018vyg,Prince:2021fdv,Rosenberg:2022sdy} yield:
\begin{eqnarray}
\rd h & = & 99.10 \pm 0.81 ~{\rm Mpc}, \nonumber \\
\Om & = & 0.3154 \pm 0.0065,
\label{eq:Planck}
\end{eqnarray}
which is in $2.1\sigma$ tension with the DESI DR2 results\footnote{To estimate the statistical significance of the discrepancy we compute the relative $\chi^2$ between the two-parameter posteriors (see Eq.~(18) of \cite{DESI:2025zgx}) and convert it into a probability-to-exceed (PTE) value. Note that our Planck dataset includes Planck PR4 CMB lensing, while \cite{DESI:2025zgx} used the combined Planck PR4 and ACT DR6 lensing likelihood that yields a $2.3\sigma$ tension.}. Similarly, high-resolution temperature and polarization data from the Atacama Cosmology Telescope 6th Data Release (ACT DR6)~\cite{ACT:2025fju,ACT:2025tim}, combined with ACT DR6 CMB lensing, yield:
\begin{eqnarray}
\rd h & = & 97.9 \pm 1.2 ~{\rm Mpc}, \nonumber \\
\Om & = & 0.3244 \pm 0.0095 ,
\label{eq:ACT}
\end{eqnarray}
in $2.6\sigma$ tension with DESI DR2. Taken together, these discrepancies between BAO and CMB-derived parameters within the $\Lambda$CDM framework may point to the need for new physics beyond the standard cosmological model.

\begin{figure}[htbp]
	\centering
	\includegraphics[width=0.45\textwidth]{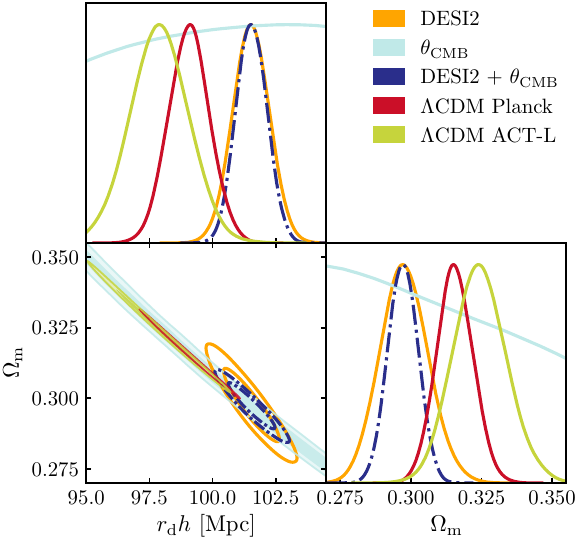}
	\caption{The 68\% and 95\% confidence level (CL) contours and posterior distributions of $\Om$ and $\rdh$ from DESI DR2 BAO (DESI2), and combination of DESI2 with the CMB acoustic scale angle $\theta_{\rm CMB}$. Contours for the $\Lambda$CDM model fit to the Planck data and to the combination of ACT DR6 CMB anisotropies and CMB lensing (ACT-L) spectra are also shown for comparison.}
	\label{fig:summary}
\end{figure}

The situation is visually summarized in Fig~\ref{fig:summary}. The figure also shows the $r_{\rm d}h$-$\Om$ posteriors obtained from the angular size of the CMB acoustic scale, $100\theta_{\rm CMB} = 1.04104 \pm 0.00025$ \cite{Tristram:2023haj}, and the combination of DESI BAO and $\theta_{\rm CMB}$. We treat $\theta_{\rm CMB}$ (also known as $\theta_\star$, see \cite{Pogosian:2024ykm} for details) as an additional transverse BAO data point at $z_\star = 1090$, assuming $r_\star \approx r_{\rm d}/1.02$ and taking $r_{\rm d}$ as a free parameter\footnote{For this, we modified {\tt CAMB}~\cite{Lewis:1999bs} to make $r_{\rm d}$ a primary parameter, as opposed to one derived from a recombination code such as {\tt RECFAST}, and sampled it as a free parameter along with $H_0$ and $\Omega_{\rm m} h^2$. More details can be found in Methods of \cite{Pogosian:2024ykm}.}. BAO and CMB measure the angular size of the same (up to a known factor) standard ruler at different redshifts, and Fig~\ref{fig:summary} shows that there is a value of $r_{\rm d}$ for which these measurements are perfectly consistent with a $\Lambda$CDM expansion history across all redshifts in the $z<1090$ range, with the CMB band passing straight through the centre of the BAO contour. Thus, the CMB-BAO tension, evident from the comparison to the Planck and ACT-L $\Lambda$CDM contours, obtained from the full CMB data and with $r_{\rm d}$ computed from a model, appears due to reasons other than the late-time expansion history. As also evident from Fig.~2 of Lynch et al \cite{Lynch:2024hzh}, {\it the discrepancy between CMB and BAO arises only when the standard recombination model is employed to compute $r_{\rm d}$.}

In addition to the above-mentioned tension with CMB, the DESI BAO-determined value of $\Om=0.2975 \pm 0.0086$ is notably lower than the values extracted from the latest compilations of uncalibrated Type Ia supernovae (SN) luminosity distances. Namely, $\Om = 0.334 \pm 0.018$ from the Pantheon Plus (PP)~\cite{Brout:2022vxf}, $\Om= 0.356 \pm 0.027$ from the Union3~\cite{Rubin:2023ovl}, and $\Om = 0.352 \pm 0.017$ from the Dark Energy Survey 5 Year (DES5Y)~\cite{DES:2024jxu} datasets, are in $1.8\sigma$, $2.1\sigma$ and $2.9\sigma$ tension with the $\Lambda$CDM-based DESI BAO value, respectively. This tension between BAO and SN has been extensively studied~\cite{Colgain:2024ksa,Colgain:2024xqj,Liu:2024gfy,Efstathiou:2024xcq,Giare:2024ocw,Colgain:2024mtg,Sousa-Neto:2025gpj,Gialamas:2024lyw,Chan-GyungPark:2024mlx,Gu:2025xie} and shown to depend substantially on the datasets and inclusion of particular data points.

Both of the above-mentioned tensions can be alleviated by allowing for dynamical dark energy (DDE)~\cite{DESI:2025zgx,DESI:2025fii,Gu:2025xie}, such as an energy density component with an equation of state given by~\cite{Chevallier:2000qy, PhysRevLett.90.091301}
\be
w(a) = w_0 + w_a(1-a),
\ee
hereafter, the $w_0w_a$ model. The DESI BAO data, by itself, prefers the $w_0w_a$ model at $1.7\sigma$~\cite{DESI:2025zgx}. Combining BAO and SN, DESI+PP, DESI+Union3 and DESI+DESY5 favour $w_0w_a$ at $1.7\sigma$, $2.7\sigma$ and $3.3\sigma$, respectively, over $\Lambda$CDM~\cite{DESI:2025zgx}. Notably, combining DESI and PP does not actually increase the preference for DDE, despite the $1.8\sigma$ tension between them in $\Lambda$CDM. When fit to the combination of Planck CMB and DESI BAO, the $w_0w_a$ model is preferred at $2.8\sigma$ level, based on the improvement in fit to the data with $\Delta \chi^2 = -10.8$ (see Table~\ref{t:params_main}). The preference for the $w_0w_a$ model increases to the $3.1\sigma$ level, when combining Planck PR4 and ACT DR6 lensing likelihood (see Table VI in~\cite{DESI:2025zgx}). Putting it all together, the combination of DESI BAO, CMB and SN yields a preference for $w_0w_a$ at $2.8-4.2\sigma$~\cite{DESI:2025zgx} (see also \cite{RoyChoudhury:2024wri,Park:2024pew,Park:2025azv,Chakraborty:2025syu,RoyChoudhury:2025dhe}), depending on the SN dataset used. If true, a detection of DDE would be a major breakthrough with profound implications for physics.

A resolution of these tensions by DDE comes with the price of having a notably lower $H_0$, thus exacerbating the Hubble tension. The latter refers to the $5.7\sigma$ discrepancy between the value of the Hubble constant, $H_0 = 67.36 \pm 0.54$ km/s/Mpc, inferred from the Planck CMB data~\cite{Planck:2018vyg} under the assumption of $\Lambda$CDM, and $H_0 = 73.17 \pm 0.86$ km/s/Mpc measured by the SH0ES collaboration using SN calibrated on Cepheid stars~\cite{Breuval:2024lsv}. (As the 2022 SH0ES value of $H_0=73.04 \pm 1.04$ km/s/Mpc~\cite{Riess:2021jrx} is still widely used in the literature, we will refer to both the 2022 and 2024 results in this paper.). Alternative SN calibration approaches also yield larger $H_0$, albeit with larger uncertainties, {\it e.g.} the CCHP collaboration reports $H_0 = 70.4 \pm 1.9$ km/s/Mpc using the Tip of the Red Giant Branch (TRGB) stars calibration method~\cite{Freedman:2024eph}. With multiple $\Lambda$CDM tensions at play, and DDE solving two of them while making another one worse, it is reasonable to ask if other extensions of $\Lambda$CDM could do better. 

In this paper, we show that a modified ionization history offers an alternative explanation for the $r_{\rm d}h$-$\Omega_m$ tension, yielding a fit to the CMB and BAO data comparable to that of DDE, while also helping relieve the Hubble tension. While DDE affects the late-time expansion, changes to early-Universe physics, particularly around the time of recombination, can also impact the inferred cosmological parameters. Such modifications may arise from physically motivated scenarios, such as the presence of primordial magnetic fields (PMF)~\cite{Jedamzik:2013gua,Jedamzik:2020krr,Thiele:2021okz,Rashkovetskyi:2021rwg,Jedamzik:2025cax}, or variations in fundamental constants~\cite{Hart:2019dxi,Sekiguchi:2020teg,Hart:2021kad,Baryakhtar:2024rky,Schoneberg:2024ynd}. Similar points were made in \cite{Lynch:2024hzh}, who fit a flexible 7-parameter modified recombination model developed in \cite{Lynch:2024gmp} to DESI DR1 BAO and Planck CMB data and compared it to that of the $w_0w_a$ model. The main novel aspects of our paper, aside from the specific focus on addressing the apparent BAO-CMB tension, is the fit to the more recent data that included DESI DR2 BAO and ACT DR6, presenting results with and without combining them with the Pantheon+ SN, and employing a simpler phenomenologically motivated 4-parameter parameterization of the ionization history. The latter has the advantage of reduced risk of over-fitting and a higher frequentist significance (FS) \cite{Wilks:1938dza, Herold:2025hkb} due to fewer parameters, while also yielding lower uncertainty in $H_0$. Related analyses, also using methods of \cite{Lynch:2024hzh} and with similar conclusions, were recently performed in \cite{ACT:2025tim}, which appeared shortly before the submission of our paper, and in the SPT-3G D1 paper~\cite{SPT-3G:2025bzu}, which appeared shortly after.

\section{Analysis details}

In our study of $\Lambda$CDM with a modified recombination history, we follow the methodology of~\cite{Mirpoorian:2024fka} (see also~\cite{Lee:2022gzh,Lynch:2024gmp} for related work). Specifically, we use a four-parameter (4-par) model for the redshift evolution of the ionized fraction $x_e(z)$, inspired by scenarios with primordial magnetic fields (PMF)~\cite{Jedamzik:2023rfd}. The model modifies the standard recombination history $x_e^{(0)}(z)$, computed for a given cosmology, by shifting it in redshift by $\Delta z_{\rm shift}$ and introducing a Gaussian-shaped bump centered at $z_{\rm b}$ with amplitude $A_{\rm b}$ and width $\sigma_{\rm b}$: $x_e(z) = x_e^{(0)}(z - \Delta z_\mathrm{shift}) (1 + A_\mathrm{b} \exp[-(z - z_\mathrm{b})^2/2 \sigma_\mathrm{b}^2])$. This parametrization captures a range of modified recombination histories that can improve the fit to the combination of Planck CMB and DESI BAO data, while accommodating significantly higher values of the Hubble constant. As shown in~\cite{Mirpoorian:2024fka}, the 4-par model performs as well as a more flexible 7-node cubic-spline parameterization similar to that used in {\tt ModRec}~\cite{Lynch:2024gmp,Lynch:2024hzh,ACT:2025tim,SPT-3G:2025bzu}, while reducing the risk of overfitting the data. Such models allow for earlier recombination, a reduced sound horizon, and hence a larger value of $H_0$~\cite{Mirpoorian:2024fka}. The net effect is an increase in $r_{\rm d}h$ and a decrease in $\Omega_m$, bringing the 4-par model into good agreement with DESI BAO observations. We now compare its performance to that of the $w_0w_a$ model.

\begin{figure*}[!htbp]
	\centering
	\begin{subfigure}
		\centering
		\includegraphics[width=0.48\textwidth]{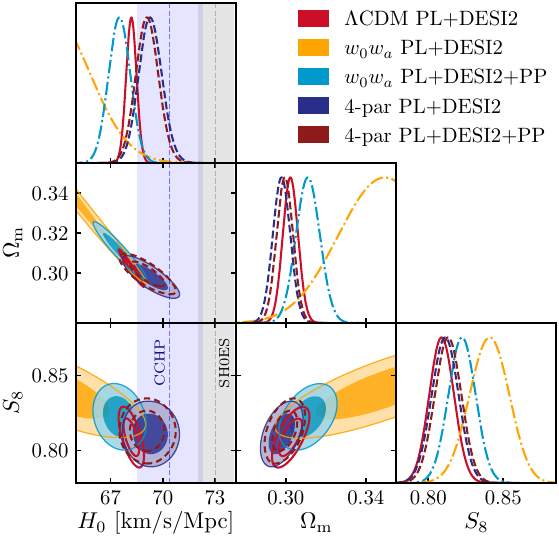}
	\end{subfigure}
	\hspace{0.1cm}
		\begin{subfigure}
		\centering
		\includegraphics[width=0.48\textwidth]{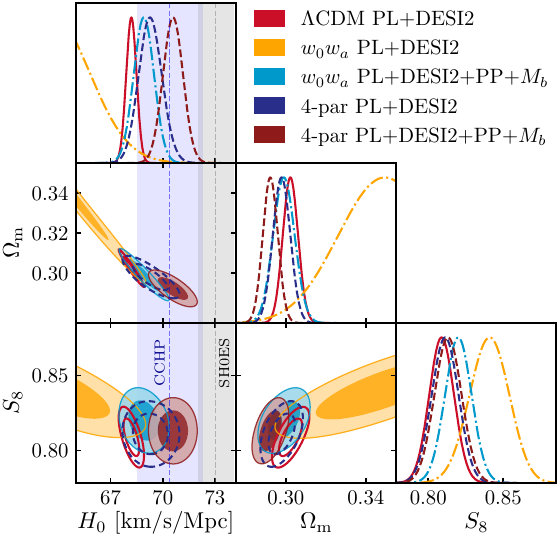}
	\end{subfigure}
	\caption{The 68\% and 95\% CL contours for the $H_0$, $\Om$, and $S_8$ along with their posterior distributions for the $\Lambda$CDM, $w_0w_a$, and 4-parameter models fit to combinations of Planck (PL), DESI2, and PP (left), and PP with the SN brightness magnitude prior from SH0ES (PP+$M_b$) (right). For reference, the blue and grey vertical bands show the 68\% CL intervals corresponding to $H_0 = 70.4 \pm 1.9$ km/s/Mpc from CCHP, and $H_0 = 73.04 \pm 1.04$ km/s/Mpc from SH0ES, respectively.}
	\label{fig:contours}
\end{figure*}

We implement the modified $x_e(z)$ in the Boltzmann code {\tt CAMB}~\cite{Lewis:1999bs}, and use {\tt Cobaya}~\cite{Torrado:2020dgo} to perform the MCMC analysis. When combining DESI DR2 BAO (DESI2) with Planck CMB data (PL), we use the combination of Planck NPIPE {\tt CamSpec} high-$\ell$ TTTEEE likelihood, Planck PR3 low-$\ell$ TT and EE, and PR4 CMB lensing~\cite{Planck:2018vyg,Prince:2021fdv,Rosenberg:2022sdy}. For combinations with ACT DR6 (parameter tables provided in the Appendix) we use the foreground-marginalized likelihood, {\tt ACT-lite}~\cite{ACT:2025fju}, combined with Planck and ACT CMB lensing (P-ACT-L). We use broad, flat, uninformative priors on the six standard cosmological parameters, while the four recombination model parameters have uniform priors $A_{\rm b} \in [-0.5, 0.5]$, $z_{\rm b} \in [500, 1500]$, $\sigma_\mathrm{b} \in [10, 500]$, and $\Delta z_{\rm shift} \in [-100, 100]$. Figure~\ref{fig:contours} shows the posterior distributions of key cosmological parameters for three models -- $\Lambda$CDM, $w_0w_a$, and the 4-par modified recombination model, fit to various combinations of datasets: PL + DESI2; PL + DESI2 + uncalibrated PP supernovae; and PL + DESI2 + PP + the 2022 SH0ES prior on the intrinsic SN brightness magnitude, $M_b = -19.253 \pm 0.027$~\cite{Riess:2021jrx}. The corresponding marginalized parameter means, uncertainties, and best-fit $\chi^2$ values are listed in Table~\ref{t:params_main}.

\section{Results}

\begin{table*}[!htbp]
	\centering
	\begin{ruledtabular}
		\begin{tabular}{|c|cccc|ccc|c|}
			{\bf Name}
			& $H_0$
			& $\Omega_{\rm m}$
			& $S_8$
			& $r_{\rm d} h$
			& $\chi^2_{\rm PL}$
			& $\chi^2_{\rm BAO}$
			& $\chi^2_{\rm SN}$
			& $\chi^2_{\rm total}$ \\[0.5ex]
			\colrule
			\rule{0pt}{3ex}
			{\boldmath $\Lambda$}{\bf CDM} PL+DESI2
			& \multirow{1}{*}{$68.19 \pm 0.28$}
			& \multirow{1}{*}{$0.302 \pm 0.004$}
			& \multirow{1}{*}{$0.809 \pm 0.008$}
			& \multirow{1}{*}{$100.78 \pm 0.47$}
			& \multirow{1}{*}{10978.0}
			& \multirow{1}{*}{11.91}
			& \multirow{1}{*}{$-$}
			& \multirow{1}{*}{10989.9} \\
			{\boldmath $\Lambda$}{\bf CDM} PL+DESI2+PP
			& \multirow{1}{*}{$68.11 \pm 0.28$}
			& \multirow{1}{*}{$0.304 \pm 0.004$}
			& \multirow{1}{*}{$0.811 \pm 0.008$}
			& \multirow{1}{*}{$100.62 \pm 0.47$}
			& \multirow{1}{*}{10975.5}
			& \multirow{1}{*}{13.53}
			& \multirow{1}{*}{1405.2}
			& \multirow{1}{*}{12394.3} \\
			{\boldmath $\Lambda$}{\bf CDM} PL+DESI2+PP+$M_b$
			& \multirow{1}{*}{$68.55 \pm 0.27$}
			& \multirow{1}{*}{$0.298 \pm 0.003$}
			& \multirow{1}{*}{$0.803 \pm 0.008$}
			& \multirow{1}{*}{$101.34 \pm 0.45$}
			& \multirow{1}{*}{10980.7}
			& \multirow{1}{*}{10.44}
			& \multirow{1}{*}{1407.4}
			& \multirow{1}{*}{12429.2} \\
			{\boldmath $\Lambda$}{\bf CDM} P-ACT-L+DESI2
			& $68.44 \pm 0.27$
			& $0.300 \pm 0.004$
			& $0.812 \pm 0.007$
			& $100.99 \pm 0.48$
			& 634.4
			& 11.58
			& $-$
			& 825.5 \\
			{\boldmath $\Lambda$}{\bf CDM} P-ACT-L+DESI2+PP
			& $68.35 \pm 0.26$
			& $0.301 \pm 0.004$
			& $0.814 \pm 0.007$
			& $100.84 \pm 0.47$
			& 633.7
			& 13.72
			& 1405.4
			& 2231.8 \\
			{\boldmath $\Lambda$}{\bf CDM} P-ACT-L+DESI2+PP+$M_b$
			& $68.73 \pm 0.26$
			& $0.297 \pm 0.003$
			& $0.807 \pm 0.007$
			& $101.48 \pm 0.45$
			& 635.8
			& 10.40
			& 1407.5
			& 2263.8 \\
			{\boldmath $w_0w_a$} PL+DESI2
			& \multirow{1}{*}{$63.9^{+1.7}_{-2.1}$}
			& \multirow{1}{*}{$0.350 \pm 0.021$}
			& \multirow{1}{*}{$0.841 \pm 0.012$}
			& \multirow{1}{*}{$94.2^{+2.6}_{-3.2}$}
			& \multirow{1}{*}{10971.9}
			& \multirow{1}{*}{7.15}
			& \multirow{1}{*}{$-$}
			& \multirow{1}{*}{10979.1} \\
			{\boldmath $w_0w_a$} PL+DESI2+PP
			& \multirow{1}{*}{$67.51 \pm 0.59$}
			& \multirow{1}{*}{$0.311 \pm 0.006$}
			& \multirow{1}{*}{$0.823 \pm 0.009$}
			& \multirow{1}{*}{$99.60 \pm 0.87$}
			& \multirow{1}{*}{10973.2}
			& \multirow{1}{*}{9.70}
			& \multirow{1}{*}{1402.8}
			& \multirow{1}{*}{12385.8} \\
			{\boldmath $w_0w_a$} PL+DESI2+PP+$M_b$
			& \multirow{1}{*}{$68.91 \pm 0.56$}
			& \multirow{1}{*}{$0.299 \pm 0.005$}
			& \multirow{1}{*}{$0.820 \pm 0.008$}
			& \multirow{1}{*}{$101.61 \pm 0.82$}
			& \multirow{1}{*}{10971.3}
			& \multirow{1}{*}{14.08}
			& \multirow{1}{*}{1403.8}
			& \multirow{1}{*}{12415.2} \\
			{\boldmath $w_0w_a$} P-ACT-L+DESI2
			& $64.10^{+1.8}_{-2.0}$
			& $0.347 \pm 0.021$
			& $0.843 \pm 0.012$
			& $94.3 \pm 2.8$
			& 634.0
			& 7.17
			& $-$
			& 817.9 \\
			{\boldmath $w_0w_a$} P-ACT-L+DESI2+PP
			& $67.64 \pm 0.60$
			& $0.310 \pm 0.006$
			& $0.825 \pm 0.008$
			& $99.65 \pm 0.89$
			& 634.5
			& 9.80
			& 1402.7
			& 2223.9 \\
			{\boldmath $w_0w_a$} P-ACT-L+DESI2+PP+$M_b$
			& $69.01 \pm 0.54$
			& $0.298 \pm 0.005$
			& $0.823 \pm 0.008$
			& $101.60 \pm 0.81$
			& 634.4
			& 13.63
			& 1405.0
			& 2251.9 \\
			{\bf 4-par} PL+DESI2
			& \multirow{1}{*}{$69.29 \pm 0.63$}
			& \multirow{1}{*}{$0.298 \pm 0.004$}
			& \multirow{1}{*}{$0.811 \pm 0.008$}
			& \multirow{1}{*}{$101.41 \pm 0.57$}
			& \multirow{1}{*}{10970.6}
			& \multirow{1}{*}{10.30}
			& \multirow{1}{*}{$-$}
			& \multirow{1}{*}{10980.9} \\
			{\bf 4-par} PL+DESI2+PP
			& \multirow{1}{*}{$69.10 \pm 0.62$}
			& \multirow{1}{*}{$0.300 \pm 0.004$}
			& \multirow{1}{*}{$0.813 \pm 0.008$}
			& \multirow{1}{*}{$101.19 \pm 0.53$}
			& \multirow{1}{*}{10970.8}
			& \multirow{1}{*}{10.44}
			& \multirow{1}{*}{1406.5}
			& \multirow{1}{*}{12387.7} \\
			{\bf 4-par} PL+DESI2+PP+$M_b$
			& \multirow{1}{*}{$70.59 \pm 0.54$}
			& \multirow{1}{*}{$0.292 \pm 0.004$}
			& \multirow{1}{*}{$0.813 \pm 0.008$}
			& \multirow{1}{*}{$102.22 \pm 0.50$}
			& \multirow{1}{*}{10974.9}
			& \multirow{1}{*}{12.13}
			& \multirow{1}{*}{1408.2}
			& \multirow{1}{*}{12404.8} \\
			{\bf 4-par} P-ACT-L+DESI2
			& $69.24^{+0.46}_{-0.41}$
			& $0.298 \pm 0.004$
			& $0.815 \pm 0.008$
			& $101.29 \pm 0.51$
			& 633.4
			& 10.33
			& $-$
			& 817.3 \\
			{\bf 4-par} P-ACT-L+DESI2+PP
			& $69.13 \pm 0.44$
			& $0.300 \pm 0.004$
			& $0.816 \pm 0.007$
			& $101.14 \pm 0.49$
			& 633.7
			& 11.67
			& 1405.9
			& 2224.4 \\
			{\bf 4-par} P-ACT-L+DESI2+PP+$M_b$
			& $69.88^{+0.36}_{-0.32}$
			& $0.294 \pm 0.003$
			& $0.813 \pm 0.007$
			& $101.83 \pm 0.47$
			& 633.0
			& 10.45
			& 1407.8
			& 2240.2
		\end{tabular}
	\end{ruledtabular}
	\caption{The mean values and the 68\% CL uncertainties for parameters of special interest, along with the best fit chi-squared values, from the MCMC runs represented in Fig.~\ref{fig:contours} as well as the runs using the combination of Planck and ACT DR6 data (P-ACT-L).}
	\label{t:params_main}
\end{table*}

When fit to PL+DESI2, the $w_0w_a$ model prefers lower values of $H_0$ and $r_{\rm d}h$, and higher values of $\Omega_m$ and $S_8$ compared to $\Lambda$CDM\footnote{$S_8 \equiv \sigma_8 \sqrt{\Omega_m /0.3}$, where $\sigma_8$ is the amplitude of matter density fluctuations smoothed over the scale of 8 Mpc/h.}. In contrast, the 4-par model favours higher $H_0$ and $r_{\rm d}h$, while maintaining $\Omega_m$ and $S_8$ values close to those in $\Lambda$CDM. Comparing the PL+DESI2 fits, the $w_0w_a$ model provides a slightly better fit, with $\Delta \chi^2 = -10.8$ relative to $\Lambda$CDM, but with a substantially lower $H_0=63.9^{+1.7}_{-2.1}$ km/s/Mpc, while the 4-par fit has $\Delta \chi^2 = -9.0$ with $H_0=69.29 \pm 0.63$ km/s/Mpc. {\it We note that these $\Delta\chi^2$ values do not account for the larger number of parameters, two for $w_0w_a$ and four for the 4-par model.}
When computing an equivalent frequentist significance (FS) for a one-dimensional Gaussian distribution (as in \cite{SPT-3G:2025bzu}), which takes into account the extra number of theory parameters, we find a $2.8\sigma$ and $1.9\sigma$ preference for the $w_0w_a$ and the 4-par models over $\Lambda$CDM, respectively.  

Adding the PP SN data substantially reduces the uncertainties in the $w_0w_a$ parameters, shifting the preferred $H_0$ and $r_{\rm d}h$ values closer to the $\Lambda$CDM values. Comparing the PL+DESI2+PP fits, the $w_0w_a$ model yields $\Delta \chi^2 = -8.5$ ($2.5\sigma$ FS) relative to $\Lambda$CDM and $H_0=67.51 \pm 0.59$ km/s/Mpc, while the 4-par model has $\Delta \chi^2 = -6.6$ ($1.4\sigma$ FS) with $H_0 = 69.10 \pm 0.62$ km/s/Mpc. Similar conclusions are drawn from the fits using the combination of Planck and ACT DR6 data (see Table \ref{t:params_act}). 

While the $w_0w_a$ model fits the PL+DESI2+PP combination slightly better, the 4-par model still provides an excellent fit overall, and to each of the datasets separately, {\it while providing a substantial relief of the Hubble tension.} In particular, the tension with the 2022 (2024) SH0ES value of $H_0=73.04 \pm 1.04$ km/s/Mpc~\cite{Riess:2021jrx} ($H_0 = 73.17 \pm 0.86$ km/s/Mpc~\cite{Breuval:2024lsv}) is reduced to $3.2\sigma$ ($3.8\sigma$) for the 4-par fit to PL+DESI2+PP, while it is $4.6\sigma$ ($5.6\sigma$) for $\Lambda$CDM and $4.6\sigma$ ($5.4\sigma$) for $w_0w_a$ ((Note that combining PL+DESI2 with Union3 or DESY5 SN instead of PP yields lower values of $H_0$~\cite{DESI:2025zgx} and a larger tension.)). For the P-ACT-L+DESI2+PP combination, the Hubble tension is $3.4\sigma$ ($4.2\sigma$), $4.4\sigma$ ($5.4\sigma$), and $4.5\sigma$ ($5.3\sigma$) for the 4-par, $\Lambda$CDM and $w_0w_a$, respectively. Note that the residual $H_0$ tension for the 4-par model is relatively high, despite the higher mean value, due to the small uncertainty. Lower levels of tension were reported from the 7-parameter {\tt ModRec}-based analysis~\cite{ACT:2025tim,SPT-3G:2025bzu} that yield similar mean values, but substantially larger uncertainties in $H_0$.

The preference for low $H_0$ and higher $S_8$ in DDE models should come as a red flag. While the most recent analysis of the DES and Kilo-Degree Survey (KiDS) data \cite{Stolzner:2025htz} showed that the $S_8$ they measure is no longer in tension with the value predicted by the Planck-best-fit LCDM, higher $S_8$ values are generally in lesser agreement with current galaxy weak lensing surveys. And the Hubble tension has not yet been resolved. Adding the 2022 SH0ES prior on $M_b$ to the fits makes the advantage of the 4-par model more apparent. The PL+DESI2+PP+$M_b$ fit of the $w_0w_a$ model has an improvement of $\Delta \chi^2 = -14$, which corresponds to a $3.3\sigma$ FS improvement over $\Lambda$CDM, but the 4-parameter model performs substantially better, with $\Delta \chi^2 = -24.4$ and a FS preference of $4\sigma$, primarily due to its prediction of a larger Hubble constant.

\section{Conclusions}

In summary, a tension exists between the values of $r_{\rm d}h$ and $\Omega_m$ inferred from recent DESI BAO observations and those derived from Planck data within the framework of $\Lambda$CDM. This discrepancy becomes even more pronounced when incorporating ACT DR6 CMB data. Notably, the tension does not arise from any fundamental disagreement between the BAO and CMB datasets themselves, as both yield consistent measurements of the angular acoustic scale and $\Omega_m$. Rather, the tension emerges when interpreting both datasets through the lens of standard $\Lambda$CDM, applied uniformly to both early- and late-Universe physics. The DESI collaboration has interpreted this tension as preference for dynamical dark energy. However, as we have shown, it may instead point to a modified recombination history. Both DDE and modified recombination models improve the overall fit to DESI and Planck data, but they differ in their impact on the Hubble tension: DDE worsens it by preferring lower $H_0$, while modified recombination partially alleviates it by allowing for a smaller sound horizon. When a global fit includes the SH0ES prior, modified recombination is favored as a resolution to the $r_{\rm d}h$-$\Omega_m$ tension. Whether this tension strengthens or fades with future data remains an open question.

While this work was in its final preparation stages, two other studies~\cite{Pang:2025lvh,Chaussidon:2025npr,Toda:2025dzd} appeared on the {\tt arXiv}, showing that there can be other ``early Universe'' solutions to the $r_{\rm d}h$-$\Omega_m$ tension that allow for higher values of $H_0$. Ultimately, the viability of any solution to the current cosmological tensions will depend on its ability to fit future data and whether it can be derived from a compelling, first-principles theoretical framework.

{\it Acknowledgments.} We thank Kazuya Koyama and Gong-Bo Zhao for useful discussions. This research was enabled in part by support provided by the BC DRI Group and the Digital Research Alliance of Canada ({\tt alliancecan.ca}). S.H.M and L.P. are supported in part by the National Sciences and Engineering Research Council (NSERC) of Canada.

{\it Data Availability:} The datasets generated and/or analyzed during the current study are available from the corresponding author on reasonable request. The observational data used (Planck, DESI, etc) are publicly available from their respective collaboration websites.

{\it Code Availability:} The code used in this study is not publicly available but may be made available from the corresponding author on reasonable request.

\appendix
\onecolumngrid
\pagebreak
\section{Parameter tables}
\label{app:params}

\begin{table*}[!htbp]
	\centering
	\begin{ruledtabular}
		\begin{tabular}{|c|cccccc|}
			\multirow{2}{*}{\bf Parameter}
			& {\boldmath $\Lambda$}{\bf CDM} PL 
			& {\boldmath $\Lambda$}{\bf CDM} PL 
			& {\boldmath $w_0w_a$} PL
			& {\boldmath $w_0w_a$} PL 
			& {\bf 4-par} PL
			& {\bf 4-par} PL \\
			& +DESI2 
			& +DESI2+PP
			& +DESI2 
			& +DESI2+PP
			& +DESI2
			& +DESI2+PP \\
			\colrule
			\rule{0pt}{3ex}
			$100~\Omega_{\rm b} h^2$
			& $2.234 \pm 0.012$
			& $2.232 \pm 0.012$
			& $2.221 \pm 0.013$
			& $2.225 \pm 0.013$
			& $2.247 \pm 0.016$
			& $2.244 \pm 0.015$
			\\[0.5ex]
			$100~\Omega_{\rm c} h^2$
			& $11.761 \pm 0.062$
			& $11.779 \pm 0.061$
			& $11.939 \pm 0.083$
			& $11.885 \pm 0.080$
			& $11.990 \pm 0.150$
			& $11.990 \pm 0.160$
			\\[0.5ex]
			$\tau_{\rm reio}$
			& $0.059 \pm 0.007$
			& $0.059 \pm 0.007$
			& $0.051 \pm 0.007$
			& $0.054 \pm 0.007$
			& $0.055 \pm 0.007$
			& $0.055 \pm 0.007$
			\\[0.5ex]
			log$(10^{10} A_{\rm s})$
			& $3.048 \pm 0.014$
			& $3.047 \pm 0.014$
			& $3.033 \pm 0.014$
			& $3.038 \pm 0.014$
			& $3.041 \pm 0.015$
			& $3.040 \pm 0.015$
			\\[0.5ex]
			$n_{\rm s}$
			& $0.969 \pm 0.003$
			& $0.968 \pm 0.003$
			& $0.965 \pm 0.004$
			& $0.966 \pm 0.004$
			& $0.965 \pm 0.008$
			& $0.965 \pm 0.007$
			\\[0.5ex]
			$H_0$\,[km/s/Mpc]
			& $68.19 \pm 0.28$
			& $68.11 \pm 0.28$
			& $63.9^{+1.7}_{-2.1}$
			& $67.51 \pm 0.59$
			& $69.29 \pm 0.63$
			& $69.10 \pm 0.62$
			\\[0.5ex]
			$\Omega_{\rm m}$
			& $0.302 \pm 0.004$
			& $0.304 \pm 0.004$
			& $0.350 \pm 0.021$
			& $0.311 \pm 0.006$
			& $0.298 \pm 0.004$
			& $0.300 \pm 0.004$
			\\[0.5ex]
			$\om$
			& $0.141 \pm 0.001$
			& $0.141 \pm 0.001$
			& $0.142 \pm 0.001$
			& $0.142 \pm 0.001$
			& $0.143 \pm 0.002$
			& $0.143 \pm 0.002$
			\\[0.5ex]
			$S_8$
			& $0.809 \pm 0.008$
			& $0.811 \pm 0.008$
			& $0.841 \pm 0.012$
			& $0.823 \pm 0.009$
			& $0.811 \pm 0.008$
			& $0.813 \pm 0.008$
			\\[0.5ex]
			$\sigma_8$
			& $0.806 \pm 0.006$
			& $0.806 \pm 0.006$
			& $0.780^{+0.016}_{-0.018}$
			& $0.808 \pm 0.009$
			& $0.814 \pm 0.008$
			& $0.814 \pm 0.008$
			\\[0.5ex]
			$r_\star$ [Mpc]
			& $145.09 \pm 0.17$
			& $145.05 \pm 0.16$
			& $144.72 \pm 0.20$
			& $144.83 \pm 0.20$
			& $143.70 \pm 0.81$
			& $143.76 \pm 0.83$
			\\[0.5ex]
			$r_{\rm d}$ [Mpc]
			& $147.78 \pm 0.19$
			& $147.75 \pm 0.18$
			& $147.44 \pm 0.21$
			& $147.54 \pm 0.21$
			& $146.37 \pm 0.80$
			& $146.44 \pm 0.82$
			\\[0.5ex]
			$r_{\rm d}h$ [Mpc]
			& $100.78 \pm 0.47$
			& $100.62 \pm 0.47$
			& $94.2^{+2.6}_{-3.2}$
			& $99.60 \pm 0.87$
			& $101.41 \pm 0.57$
			& $101.19 \pm 0.53$
			\\[0.5ex]
			$z_\star$
			& $1089.77 \pm 0.17$
			& $1089.81 \pm 0.17$
			& $1090.09 \pm 0.20$
			& $1089.99 \pm 0.20$
			& $1098.00 \pm 5.20$
			& $1097.30 \pm 5.20$
			\\[0.5ex]
			$z_{\rm d}$
			& $1059.69 \pm 0.27$
			& $1059.67 \pm 0.27$
			& $1059.53 \pm 0.28$
			& $1059.59 \pm 0.28$
			& $1067.50 \pm 4.70$
			& $1066.90 \pm 4.70$
			\\[0.5ex]
			$w_0$
			& $-$
			& $-$
			& $-0.45 \pm 0.21$
			& $-0.842 \pm 0.054$
			& $-$
			& $-$
			\\[0.5ex]
			$w_a$
			& $-$
			& $-$
			& $-1.64 \pm 0.59$
			& $-0.59^{+0.22}_{-0.18}$
			& $-$
			& $-$
			\\[0.5ex]
			$A_{\rm b}$
			& $-$
			& $-$
			& $-$
			& $-$
			& $< 0.215$
			& $< 0.208$
			\\[0.5ex]
			$z_{\rm b}$
			& $-$
			& $-$
			& $-$
			& $-$
			& $926^{+51}_{-34}$
			& $930^{+41}_{-36}$
			\\[0.5ex]
			$\sigma_{\rm b}$
			& $-$
			& $-$
			& $-$
			& $-$
			& $163^{+60}_{-40}$
			& $169^{+60}_{-30}$
			\\[0.5ex]
			$\Delta z_{\rm shift}$
			& $-$
			& $-$
			& $-$
			& $-$
			& $-20.0^{+11}_{-7.7}$
			& $-19.4^{+11}_{-6.8}$
			\\[0.5ex]
			\colrule
			\rule{0pt}{3ex}
			$\chi^2_{\rm PL}$
			& 10977.96
			& 10975.51
			& 10971.90
			& 10973.23
			& 10970.63
			& 10970.76
			\\[0.5ex]
			$\chi^2_{\rm BAO}$
			& 11.91
			& 13.53
			& 7.15
			& 9.70
			& 10.30
			& 10.44
			\\[0.5ex]
			$\chi^2_{\rm SN}$
			& $-$
			& 1405.25
			& $-$
			& 1402.82
			& $-$
			& 1406.47
			\\[0.5ex]
			\colrule
			\rule{0pt}{3ex}
			$\chi^2_{\rm total}$
			& 10989.87
			& 12394.29
			& 10979.05
			& 12385.76
			& 10980.93
			& 12387.67
		\end{tabular}
	\end{ruledtabular}
\caption{Mean values and the 68\% CL intervals of the parameters and the $\chi^2_{\rm BF}$ values for the $\Lambda$CDM, $w_0w_a$, and 4-parameter models fit to combination of Planck, DESI2, and PP.}
\label{t:params_pl}
\end{table*}

\begin{table*}[!htbp]
	\begin{ruledtabular}
		\begin{tabular}{|c|cccccc|}
			\multirow{2}{*}{\bf Parameter}
			& {\boldmath $\Lambda$}{\bf CDM} P-ACT-L
			& {\boldmath $\Lambda$}{\bf CDM} P-ACT-L
			& {\boldmath $w_0w_a$} P-ACT-L
			& {\boldmath $w_0w_a$} P-ACT-L
			& {\bf 4-par} P-ACT-L
			& {\bf 4-par} P-ACT-L \\[0.5ex]
			& +DESI2
			& +DESI2+PP
			& +DESI2
			& +DESI2+PP
			& +DESI2
			& +DESI2+PP \\[0.5ex]
			\colrule
			\rule{0pt}{3ex}
			$100~\Omega_{\rm b} h^2$
			& $2.258 \pm 0.010$
			& $2.256 \pm 0.010$
			& $2.250 \pm 0.010$
			& $2.252 \pm 0.011$
			& $2.246 \pm 0.013$
			& $2.245 \pm 0.013$ \\[0.5ex]
			$100~\Omega_{\rm c} h^2$
			& $11.74 \pm 0.064$
			& $11.76 \pm 0.063$
			& $11.90 \pm 0.084$
			& $11.85 \pm 0.085$
			& $12.00 \pm 0.120$
			& $12.01 \pm 0.120$
			\\[0.5ex]
			$\tau_{\rm reio}$
			& $0.064^{+0.006}_{-0.007}$
			& $0.064^{+0.006}_{-0.007}$
			& $0.059^{+0.005}_{-0.006}$
			& $0.061 \pm 0.006$
			& $0.059^{+0.005}_{-0.006}$
			& $0.059^{+0.005}_{-0.006}$ \\[0.5ex]
			log$(10^{10} A_{\rm s})$
			& $3.062^{+0.011}_{-0.012}$
			& $3.061^{+0.010}_{-0.012}$
			& $3.050 \pm 0.011$
			& $3.055^{+0.010}_{-0.012}$
			& $3.050 \pm 0.011$
			& $3.050^{+0.010}_{-0.012}$
			\\[0.5ex]
			$n_{\rm s}$
			& $0.975 \pm 0.003$
			& $0.975 \pm 0.003$
			& $0.972 \pm 0.003$
			& $0.973 \pm 0.003$
			& $0.961 \pm 0.006$
			& $0.961 \pm 0.006$ \\[0.5ex]
			$H_0$\,[km/s/Mpc]
			& $68.44 \pm 0.27$
			& $68.35 \pm 0.26$
			& $64.10^{+1.8}_{-2.0}$
			& $67.64 \pm 0.60$
			& $69.24^{+0.46}_{-0.41}$
			& $69.13 \pm 0.44$ \\[0.5ex]
			$\Omega_{\rm m}$
			& $0.300 \pm 0.004$
			& $0.301 \pm 0.004$
			& $0.347 \pm 0.021$
			& $0.310 \pm 0.006$
			& $0.298 \pm 0.004$
			& $0.300 \pm 0.004$ \\[0.5ex]
			$\Omega_{\rm m} h^2$
			& $0.141 \pm 0.001$
			& $0.141 \pm 0.001$
			& $0.142 \pm 0.001$
			& $0.142 \pm 0.001$
			& $0.143 \pm 0.001$
			& $0.143 \pm 0.001$ \\[0.5ex]
			$S_8$
			& $0.812 \pm 0.007$
			& $0.814 \pm 0.007$
			& $0.843 \pm 0.012$
			& $0.825 \pm 0.008$
			& $0.815 \pm 0.008$
			& $0.816 \pm 0.007$ \\[0.5ex]
			$\sigma_8$
			& $0.812 \pm 0.005$
			& $0.812 \pm 0.005$
			& $0.784 \pm 0.016$
			& $0.812 \pm 0.008$
			& $0.817 \pm 0.005$
			& $0.817 \pm 0.005$ \\[0.5ex]
			$r_\star$ [Mpc]
			& $144.96 \pm 0.17$
			& $144.92 \pm 0.17$
			& $144.60 \pm 0.21$
			& $144.72 \pm 0.21$
			& $143.63^{+0.50}_{-0.60}$
			& $143.63^{+0.49}_{-0.59}$
			\\[0.5ex]
			$r_{\rm d}$ [Mpc]
			& $147.57 \pm 0.19$
			& $147.53 \pm 0.19$
			& $147.22 \pm 0.22$
			& $147.33 \pm 0.23$
			& $146.29^{+0.49}_{-0.59}$
			& $146.30^{+0.48}_{-0.58}$ \\[0.5ex]
			$r_{\rm d} h$ [Mpc]
			& $100.99 \pm 0.48$
			& $100.84 \pm 0.47$
			& $94.3 \pm 2.8$
			& $99.65 \pm 0.89$
			& $101.29 \pm 0.51$
			& $101.14 \pm 0.49$ \\[0.5ex]
			$z_\star$
			& $1089.40^{+0.14}_{-0.16}$
			& $1089.44 \pm 0.15$
			& $1089.65 \pm 0.17$
			& $1089.57 \pm 0.17$
			& $1098.6^{+4.0}_{-2.9}$
			& $1098.4^{+3.8}_{-3.2}$ \\[0.5ex]
			$z_{\rm d}$
			& $1060.21 \pm 0.23$
			& $1060.19 \pm 0.23$
			& $1060.16 \pm 0.23$
			& $1060.17 \pm 0.23$
			& $1068.3^{+3.6}_{-2.6}$
			& $1068.0^{+3.4}_{-2.8}$ \\[0.5ex]
			$w_0$
			& $-$
			& $-$
			& $-0.47 \pm 0.21$
			& $-0.849 \pm 0.055$
			& $-$
			& $-$ \\[0.5ex]
			$w_a$
			& $-$
			& $-$
			& $-1.536^{+0.62}_{-0.55}$
			& $-0.53^{+0.21}_{-0.19}$
			& $-$
			& $-$ \\[0.5ex]
			$A_{\rm b}$
			& $-$
			& $-$
			& $-$
			& $-$
			& $< 0.123$
			& $< 0.134$ \\[0.5ex]
			$z_{\rm b}$
			& $-$
			& $-$
			& $-$
			& $-$
			& $< 802$
			& $< 792$ \\[0.5ex]
			$\sigma_{\rm b}$
			& $-$
			& $-$
			& $-$
			& $-$
			& $< 259$
			& $< 264$ \\[0.5ex]
			$\Delta z_{\rm shift}$
			& $-$
			& $-$
			& $-$
			& $-$
			& $-12.7^{+5.0}_{-4.4}$
			& $-12.4^{+5.6}_{-4.2}$ \\[0.5ex]
			\colrule
			\rule{0pt}{3ex}
			$\chi^2_{\rm PL}$
			& 634.40
			& 633.72
			& 634.00
			& 634.47
			& 633.39
			& 633.73 \\[0.5ex]
			$\chi^2_{\rm BAO}$
			& 11.58
			& 13.72
			& 7.17
			& 9.80
			& 10.33
			& 11.67 \\[0.5ex]
			$\chi^2_{\rm ACT-L}$
			& 179.54
			& 178.97
			& 176.71
			& 176.86
			& 173.60
			& 173.14 \\[0.5ex]
			$\chi^2_{\rm SN}$
			& $-$
			& 1405.42
			& $-$
			& 1402.75
			& $-$
			& 1405.88 \\[0.5ex]
			\colrule
			\rule{0pt}{3ex}
			$\chi^2_{\rm total}$
			& 825.52
			& 2231.83
			& 817.88
			& 2223.87
			& 817.32
			& 2224.42
		\end{tabular}
	\end{ruledtabular}
	\caption{Mean values and the 68\% CL intervals of the parameters and the $\chi^2_{\rm BF}$ values for the $\Lambda$CDM, $w_0w_a$, and 4-parameter models fit to combination of P-ACT-L, DESI2, and PP. Here, P-ACT-L refers to ACT DR6 combined with ACT DR6 CMB lensing and Planck PR3, where Planck data is cut at $\ell < 1000$ in temperature, and $\ell < 600$ in polarization, consistent with \cite{ACT:2025fju}.}
	\label{t:params_act}
\end{table*}

\begin{table*}[!htbp]
	\begin{ruledtabular}
		\begin{tabular}{|c|cccccc|}
			\multirow{3}{*}{\bf Parameter}
			& {\boldmath $\Lambda$}{\bf CDM} PL
			& {\boldmath $\Lambda$}{\bf CDM} P-ACT-L
			& {\boldmath $w_0w_a$} PL
			& {\boldmath $w_0w_a$} P-ACT-L
			& {\bf 4-par} PL
			& {\bf 4-par} P-ACT-L \\[0.5ex]
			& +DESI2
			& +DESI2
			& +DESI2
			& +DESI2
			& +DESI2
			& +DESI2 \\[0.5ex]
			& +PP+$M_b$
			& +PP+$M_b$
			& +PP+$M_b$
			& +PP+$M_b$
			& +PP+$M_b$
			& +PP+$M_b$
			\\[0.5ex]
			\colrule
			\rule{0pt}{3ex}
			$100~\Omega_{\rm b} h^2$
			& $2.244 \pm 0.012$
			& $2.264 \pm 0.010$
			& $2.228 \pm 0.012$
			& $2.255 \pm 0.011$
			& $2.256 \pm 0.016$
			& $2.256^{+0.014}_{-0.016}$
			\\[0.5ex]
			$100~\Omega_{\rm c} h^2$
			& $11.696 \pm 0.059$
			& $11.681 \pm 0.061$
			& $11.907 \pm 0.079$
			& $11.881 \pm 0.083$
			& $12.240 \pm 0.150$
			& $12.060^{+0.090}_{-0.076}$
			\\[0.5ex]
			$\tau_{\rm reio}$
			& $0.062^{+0.007}_{-0.008}$
			& $0.066^{+0.006}_{-0.007}$
			& $0.052 \pm 0.007$
			& $0.060^{+0.005}_{-0.006}$
			& $0.051 \pm 0.007$
			& $0.059^{+0.005}_{-0.006}$
			\\[0.5ex]
			log$(10^{10} A_{\rm s})$
			& $3.052 \pm 0.014$
			& $3.066 \pm 0.012$
			& $3.036 \pm 0.014$
			& $3.052^{+0.010}_{-0.011}$
			& $3.034 \pm 0.015$
			& $3.049 \pm 0.011$
			\\[0.5ex]
			$n_{\rm s}$
			& $0.971 \pm 0.003$
			& $0.976 \pm 0.003$
			& $0.965 \pm 0.004$
			& $0.972 \pm 0.003$
			& $0.959 \pm 0.007$
			& $0.960 \pm 0.006$
			\\[0.5ex]
			$H_0$\,[km/s/Mpc]
			& $68.55 \pm 0.27$
			& $68.73 \pm 0.26$
			& $68.91 \pm 0.56$
			& $69.01 \pm 0.54$
			& $70.59 \pm 0.54$
			& $69.88^{+0.36}_{-0.32}$
			\\[0.5ex]
			$\Omega_{\rm m}$
			& $0.298 \pm 0.003$
			& $0.297 \pm 0.003$
			& $0.299 \pm 0.005$
			& $0.298 \pm 0.005$
			& $0.292 \pm 0.004$
			& $0.294 \pm 0.003$
			\\[0.5ex]
			$\Omega_{\rm m} h^2$
			& $0.140 \pm 0.001$
			& $0.140 \pm 0.001$
			& $0.142 \pm 0.001$
			& $0.142 \pm 0.001$
			& $0.146 \pm 0.001$
			& $0.144 \pm 0.001$
			\\[0.5ex]
			$S_8$
			& $0.803 \pm 0.008$
			& $0.807 \pm 0.007$
			& $0.820 \pm 0.008$
			& $0.823 \pm 0.008$
			& $0.813 \pm 0.008$
			& $0.813 \pm 0.007$
			\\[0.5ex]
			$\sigma_8$
			& $0.806 \pm 0.006$
			& $0.811 \pm 0.005$
			& $0.821 \pm 0.008$
			& $0.826 \pm 0.008$
			& $0.824 \pm 0.008$
			& $0.821 \pm 0.005$
			\\[0.5ex]
			$r_\star$ [Mpc]
			& $145.18 \pm 0.16$
			& $145.07 \pm 0.17$
			& $144.75 \pm 0.19$
			& $144.62 \pm 0.21$
			& $142.15 \pm 0.74$
			& $143.08^{+0.25}_{-0.41}$
			\\[0.5ex]
			$r_{\rm d}$ [Mpc]
			& $147.84 \pm 0.18$
			& $147.65 \pm 0.19$
			& $147.45 \pm 0.21$
			& $147.22 \pm 0.22$
			& $144.83 \pm 0.73$
			& $145.72^{+0.29}_{-0.41}$
			\\[0.5ex]
			$r_{\rm d} h$ [Mpc]
			& $101.34 \pm 0.45$
			& $101.48 \pm 0.45$
			& $101.61 \pm 0.82$
			& $101.60 \pm 0.81$
			& $102.22 \pm 0.50$
			& $101.83 \pm 0.47$
			\\[0.5ex]
			$z_\star$
			& $1089.59 \pm 0.16$
			& $1089.27 \pm 0.15$
			& $1089.98 \pm 0.19$
			& $1089.57 \pm 0.18$
			& $1108.00 \pm 4.70$
			& $1102.10^{+2.3}_{-1.1}$
			\\[0.5ex]
			$z_{\rm d}$
			& $1059.88 \pm 0.27$
			& $1060.31 \pm 0.23$
			& $1059.66 \pm 0.27$
			& $1060.25 \pm 0.23$
			& $1076.80 \pm 4.20$
			& $1071.7^{+2.1}_{-0.89}$
			\\[0.5ex]
			$w_0$
			& $-$
			& $-$
			& $-0.872 \pm 0.056$
			& $-0.878 \pm 0.055$
			& $-$
			& $-$
			\\[0.5ex]
			$w_a$
			& $-$
			& $-$
			& $-0.68^{+0.23}_{-0.21}$
			& $-0.62^{+0.22}_{-0.20}$
			& $-$
			& $-$
			\\[0.5ex]
			$A_{\rm b}$
			& $-$
			& $-$
			& $-$
			& $-$
			& $0.253^{+0.084}_{-0.15}$
			& $< 0.129$
			\\[0.5ex]
			$z_{\rm b}$
			& $-$
			& $-$
			& $-$
			& $-$
			& $926 \pm 33$
			& $805^{+200}_{-50}$
			\\[0.5ex]
			$\sigma_{\rm b}$
			& $-$
			& $-$
			& $-$
			& $-$
			& $159^{+50}_{-20}$
			& $207 \pm 80$
			\\[0.5ex]
			$\Delta z_{\rm shift}$
			& $-$
			& $-$
			& $-$
			& $-$
			& $-33.7 \pm 8.6$
			& $-19.4^{+5.7}_{-3.4}$
			\\[0.5ex]
			\colrule
			\rule{0pt}{3ex}
			$\chi^2_{\rm PL}$
			& 10980.69
			& 635.84
			& 10971.31
			& 634.42
			& 10974.87
			& 633.01
			\\[0.5ex]
			$\chi^2_{\rm BAO}$
			& 10.44
			& 10.40
			& 14.08
			& 13.63
			& 12.13
			& 10.45
			\\[0.5ex]
			$\chi^2_{\rm ACT-L}$
			& $-$
			& 180.45
			& $-$
			& 176.01
			& $-$
			& 173.87
			\\[0.5ex]
			$\chi^2_{\rm SN}$
			& 1407.43
			& 1407.52
			& 1403.84
			& 1405.04
			& 1408.21
			& 1407.84
			\\[0.5ex]
			$\chi^2_{\rm SH0ES}$
			& 30.67
			& 29.55
			& 26.00
			& 22.85
			& 9.55
			& 15.06
			\\[0.5ex]
			\colrule
			\rule{0pt}{3ex}
			$\chi^2_{\rm total}$
			& 12429.24
			& 2263.77
			& 12415.24
			& 2251.95
			& 12404.76
			& 2240.24
			\\[0.5ex]
		\end{tabular}
	\end{ruledtabular}
	\caption{Mean values and the 68\% CL intervals of the parameters and the $\chi^2_{\rm BF}$ values for the $\Lambda$CDM, $w_0w_a$, and 4-parameter models fit to combination of Planck or P-ACT-L with DESI2, and PP+$M_b$.}
	\label{t:params_ppmb}
\end{table*}

\bibliographystyle{apsrev4-2}

%

\end{document}